\documentclass[12pt,prd,tightenlines,nofootinbib,showpacs,showkeys]{revtex4}
\newcommand{\be}{\begin{equation}}
\newcommand{\ee}{\end{equation}}
\usepackage{bm}
\usepackage{graphics}
\usepackage{rotating}
\usepackage{epsfig}

\begin{document}

\begin{flushright}
{\bf Preprint SSU-HEP-05/09\\
Samara State University}
\end{flushright}

\vspace{5mm}

\title{Proton polarizability effect
in the Lamb shift of the hydrogen atom}

\author{\firstname{A.P.} \surname{Martynenko}}
\email{mart@ssu.samara.ru(A.P.Martynenko)}
\affiliation{Samara State University, 443011, Pavlov Street 1, Samara, Russia}

\begin{abstract}
The proton polarizability correction to the Lamb shift of
electronic and muonic hydrogen is calculated on the basis of
isobar model and experimental data on the structure functions of
deep inelastic lepton-nucleon scattering. The contributions of
the Born terms, vector-meson exchanges and nucleon resonances are
taken into account in the construction of the photoabsorption
cross sections for transversely and longitudinally polarized
virtual photons $\sigma_{T,L}$.
\end{abstract}



\maketitle

\immediate\write16{<<WARNING: LINEDRAW macros work with emTeX-dvivers
                    and other drivers supporting emTeX \special's
                    (dviscr, dvihplj, dvidot, dvips, dviwin, etc.) >>}

\newdimen\Lengthunit       \Lengthunit  = 1.5cm
\newcount\Nhalfperiods     \Nhalfperiods= 9
\newcount\magnitude        \magnitude = 1000

\catcode`\*=11
\newdimen\L*   \newdimen\d*   \newdimen\d**
\newdimen\dm*  \newdimen\dd*  \newdimen\dt*
\newdimen\a*   \newdimen\b*   \newdimen\c*
\newdimen\a**  \newdimen\b**
\newdimen\xL*  \newdimen\yL*
\newdimen\rx*  \newdimen\ry*
\newdimen\tmp* \newdimen\linwid*

\newcount\k*   \newcount\l*   \newcount\m*
\newcount\k**  \newcount\l**  \newcount\m**
\newcount\n*   \newcount\dn*  \newcount\r*
\newcount\N*   \newcount\*one \newcount\*two  \*one=1 \*two=2
\newcount\*ths \*ths=1000
\newcount\angle*  \newcount\q*  \newcount\q**
\newcount\angle** \angle**=0
\newcount\sc*     \sc*=0

\newtoks\cos*  \cos*={1}
\newtoks\sin*  \sin*={0}

\catcode`\[=13

\def\rotate(#1){\advance\angle**#1\angle*=\angle**
\q**=\angle*\ifnum\q**<0\q**=-\q**\fi
\ifnum\q**>360\q*=\angle*\divide\q*360\multiply\q*360\advance\angle*-\q*\fi
\ifnum\angle*<0\advance\angle*360\fi\q**=\angle*\divide\q**90\q**=\q**
\def\sgcos*{+}\def\sgsin*{+}\relax
\ifcase\q**\or
 \def\sgcos*{-}\def\sgsin*{+}\or
 \def\sgcos*{-}\def\sgsin*{-}\or
 \def\sgcos*{+}\def\sgsin*{-}\else\fi
\q*=\q**
\multiply\q*90\advance\angle*-\q*
\ifnum\angle*>45\sc*=1\angle*=-\angle*\advance\angle*90\else\sc*=0\fi
\def[##1,##2]{\ifnum\sc*=0\relax
\edef\cs*{\sgcos*.##1}\edef\sn*{\sgsin*.##2}\ifcase\q**\or
 \edef\cs*{\sgcos*.##2}\edef\sn*{\sgsin*.##1}\or
 \edef\cs*{\sgcos*.##1}\edef\sn*{\sgsin*.##2}\or
 \edef\cs*{\sgcos*.##2}\edef\sn*{\sgsin*.##1}\else\fi\else
\edef\cs*{\sgcos*.##2}\edef\sn*{\sgsin*.##1}\ifcase\q**\or
 \edef\cs*{\sgcos*.##1}\edef\sn*{\sgsin*.##2}\or
 \edef\cs*{\sgcos*.##2}\edef\sn*{\sgsin*.##1}\or
 \edef\cs*{\sgcos*.##1}\edef\sn*{\sgsin*.##2}\else\fi\fi
\cos*={\cs*}\sin*={\sn*}\global\edef\gcos*{\cs*}\global\edef\gsin*{\sn*}}\relax
\ifcase\angle*[9999,0]\or
[999,017]\or[999,034]\or[998,052]\or[997,069]\or[996,087]\or
[994,104]\or[992,121]\or[990,139]\or[987,156]\or[984,173]\or
[981,190]\or[978,207]\or[974,224]\or[970,241]\or[965,258]\or
[961,275]\or[956,292]\or[951,309]\or[945,325]\or[939,342]\or
[933,358]\or[927,374]\or[920,390]\or[913,406]\or[906,422]\or
[898,438]\or[891,453]\or[882,469]\or[874,484]\or[866,499]\or
[857,515]\or[848,529]\or[838,544]\or[829,559]\or[819,573]\or
[809,587]\or[798,601]\or[788,615]\or[777,629]\or[766,642]\or
[754,656]\or[743,669]\or[731,681]\or[719,694]\or[707,707]\or
\else[9999,0]\fi}

\catcode`\[=12

\def\GRAPH(hsize=#1)#2{\hbox to #1\Lengthunit{#2\hss}}

\def\Linewidth#1{\global\linwid*=#1\relax
\global\divide\linwid*10\global\multiply\linwid*\mag
\global\divide\linwid*100\special{em:linewidth \the\linwid*}}

\Linewidth{.4pt}
\def\sm*{\special{em:moveto}}
\def\sl*{\special{em:lineto}}
\let\moveto=\sm*
\let\lineto=\sl*
\newbox\spm*   \newbox\spl*
\setbox\spm*\hbox{\sm*}
\setbox\spl*\hbox{\sl*}

\def\mov#1(#2,#3)#4{\rlap{\L*=#1\Lengthunit
\xL*=#2\L* \yL*=#3\L*
\xL*=\xscale\xL* \yL*=\yscale\yL*
\rx* \the\cos*\xL* \tmp* \the\sin*\yL* \advance\rx*-\tmp*
\ry* \the\cos*\yL* \tmp* \the\sin*\xL* \advance\ry*\tmp*
\kern\rx*\raise\ry*\hbox{#4}}}

\def\rmov*(#1,#2)#3{\rlap{\xL*=#1\yL*=#2\relax
\rx* \the\cos*\xL* \tmp* \the\sin*\yL* \advance\rx*-\tmp*
\ry* \the\cos*\yL* \tmp* \the\sin*\xL* \advance\ry*\tmp*
\kern\rx*\raise\ry*\hbox{#3}}}

\def\lin#1(#2,#3){\rlap{\sm*\mov#1(#2,#3){\sl*}}}

\def\arr*(#1,#2,#3){\rmov*(#1\dd*,#1\dt*){\sm*
\rmov*(#2\dd*,#2\dt*){\rmov*(#3\dt*,-#3\dd*){\sl*}}\sm*
\rmov*(#2\dd*,#2\dt*){\rmov*(-#3\dt*,#3\dd*){\sl*}}}}

\def\arrow#1(#2,#3){\rlap{\lin#1(#2,#3)\mov#1(#2,#3){\relax
\d**=-.012\Lengthunit\dd*=#2\d**\dt*=#3\d**
\arr*(1,10,4)\arr*(3,8,4)\arr*(4.8,4.2,3)}}}

\def\arrlin#1(#2,#3){\rlap{\L*=#1\Lengthunit\L*=.5\L*
\lin#1(#2,#3)\rmov*(#2\L*,#3\L*){\arrow.1(#2,#3)}}}

\def\dasharrow#1(#2,#3){\rlap{{\Lengthunit=0.9\Lengthunit
\dashlin#1(#2,#3)\mov#1(#2,#3){\sm*}}\mov#1(#2,#3){\sl*
\d**=-.012\Lengthunit\dd*=#2\d**\dt*=#3\d**
\arr*(1,10,4)\arr*(3,8,4)\arr*(4.8,4.2,3)}}}

\def\clap#1{\hbox to 0pt{\hss #1\hss}}

\def\ind(#1,#2)#3{\rlap{\L*=.1\Lengthunit
\xL*=#1\L* \yL*=#2\L*
\rx* \the\cos*\xL* \tmp* \the\sin*\yL* \advance\rx*-\tmp*
\ry* \the\cos*\yL* \tmp* \the\sin*\xL* \advance\ry*\tmp*
\kern\rx*\raise\ry*\hbox{\lower2pt\clap{$#3$}}}}

\def\sh*(#1,#2)#3{\rlap{\dm*=\the\n*\d**
\xL*=\xscale\dm* \yL*=\yscale\dm* \xL*=#1\xL* \yL*=#2\yL*
\rx* \the\cos*\xL* \tmp* \the\sin*\yL* \advance\rx*-\tmp*
\ry* \the\cos*\yL* \tmp* \the\sin*\xL* \advance\ry*\tmp*
\kern\rx*\raise\ry*\hbox{#3}}}

\def\calcnum*#1(#2,#3){\a*=1000sp\b*=1000sp\a*=#2\a*\b*=#3\b*
\ifdim\a*<0pt\a*-\a*\fi\ifdim\b*<0pt\b*-\b*\fi
\ifdim\a*>\b*\c*=.96\a*\advance\c*.4\b*
\else\c*=.96\b*\advance\c*.4\a*\fi
\k*\a*\multiply\k*\k*\l*\b*\multiply\l*\l*
\m*\k*\advance\m*\l*\n*\c*\r*\n*\multiply\n*\n*
\dn*\m*\advance\dn*-\n*\divide\dn*2\divide\dn*\r*
\advance\r*\dn*
\c*=\the\Nhalfperiods5sp\c*=#1\c*\ifdim\c*<0pt\c*-\c*\fi
\multiply\c*\r*\N*\c*\divide\N*10000}

\def\dashlin#1(#2,#3){\rlap{\calcnum*#1(#2,#3)\relax
\d**=#1\Lengthunit\ifdim\d**<0pt\d**-\d**\fi
\divide\N*2\multiply\N*2\advance\N*\*one
\divide\d**\N*\sm*\n*\*one\sh*(#2,#3){\sl*}\loop
\advance\n*\*one\sh*(#2,#3){\sm*}\advance\n*\*one
\sh*(#2,#3){\sl*}\ifnum\n*<\N*\repeat}}

\def\dashdotlin#1(#2,#3){\rlap{\calcnum*#1(#2,#3)\relax
\d**=#1\Lengthunit\ifdim\d**<0pt\d**-\d**\fi
\divide\N*2\multiply\N*2\advance\N*1\multiply\N*2\relax
\divide\d**\N*\sm*\n*\*two\sh*(#2,#3){\sl*}\loop
\advance\n*\*one\sh*(#2,#3){\kern-1.48pt\lower.5pt\hbox{\rm.}}\relax
\advance\n*\*one\sh*(#2,#3){\sm*}\advance\n*\*two
\sh*(#2,#3){\sl*}\ifnum\n*<\N*\repeat}}

\def\shl*(#1,#2)#3{\kern#1#3\lower#2#3\hbox{\unhcopy\spl*}}

\def\trianglin#1(#2,#3){\rlap{\toks0={#2}\toks1={#3}\calcnum*#1(#2,#3)\relax
\dd*=.57\Lengthunit\dd*=#1\dd*\divide\dd*\N*
\divide\dd*\*ths \multiply\dd*\magnitude
\d**=#1\Lengthunit\ifdim\d**<0pt\d**-\d**\fi
\multiply\N*2\divide\d**\N*\sm*\n*\*one\loop
\shl**{\dd*}\dd*-\dd*\advance\n*2\relax
\ifnum\n*<\N*\repeat\n*\N*\shl**{0pt}}}

\def\wavelin#1(#2,#3){\rlap{\toks0={#2}\toks1={#3}\calcnum*#1(#2,#3)\relax
\dd*=.23\Lengthunit\dd*=#1\dd*\divide\dd*\N*
\divide\dd*\*ths \multiply\dd*\magnitude
\d**=#1\Lengthunit\ifdim\d**<0pt\d**-\d**\fi
\multiply\N*4\divide\d**\N*\sm*\n*\*one\loop
\shl**{\dd*}\dt*=1.3\dd*\advance\n*\*one
\shl**{\dt*}\advance\n*\*one
\shl**{\dd*}\advance\n*\*two
\dd*-\dd*\ifnum\n*<\N*\repeat\n*\N*\shl**{0pt}}}

\def\w*lin(#1,#2){\rlap{\toks0={#1}\toks1={#2}\d**=\Lengthunit\dd*=-.12\d**
\divide\dd*\*ths \multiply\dd*\magnitude
\N*8\divide\d**\N*\sm*\n*\*one\loop
\shl**{\dd*}\dt*=1.3\dd*\advance\n*\*one
\shl**{\dt*}\advance\n*\*one
\shl**{\dd*}\advance\n*\*one
\shl**{0pt}\dd*-\dd*\advance\n*1\ifnum\n*<\N*\repeat}}

\def\l*arc(#1,#2)[#3][#4]{\rlap{\toks0={#1}\toks1={#2}\d**=\Lengthunit
\dd*=#3.037\d**\dd*=#4\dd*\dt*=#3.049\d**\dt*=#4\dt*\ifdim\d**>10mm\relax
\d**=.25\d**\n*\*one\shl**{-\dd*}\n*\*two\shl**{-\dt*}\n*3\relax
\shl**{-\dd*}\n*4\relax\shl**{0pt}\else
\ifdim\d**>5mm\d**=.5\d**\n*\*one\shl**{-\dt*}\n*\*two
\shl**{0pt}\else\n*\*one\shl**{0pt}\fi\fi}}

\def\d*arc(#1,#2)[#3][#4]{\rlap{\toks0={#1}\toks1={#2}\d**=\Lengthunit
\dd*=#3.037\d**\dd*=#4\dd*\d**=.25\d**\sm*\n*\*one\shl**{-\dd*}\relax
\n*3\relax\sh*(#1,#2){\xL*=\xscale\dd*\yL*=\yscale\dd*
\kern#2\xL*\lower#1\yL*\hbox{\sm*}}\n*4\relax\shl**{0pt}}}

\def\shl**#1{\c*=\the\n*\d**\d*=#1\relax
\a*=\the\toks0\c*\b*=\the\toks1\d*\advance\a*-\b*
\b*=\the\toks1\c*\d*=\the\toks0\d*\advance\b*\d*
\a*=\xscale\a*\b*=\yscale\b*
\rx* \the\cos*\a* \tmp* \the\sin*\b* \advance\rx*-\tmp*
\ry* \the\cos*\b* \tmp* \the\sin*\a* \advance\ry*\tmp*
\raise\ry*\rlap{\kern\rx*\unhcopy\spl*}}

\def\wlin*#1(#2,#3)[#4]{\rlap{\toks0={#2}\toks1={#3}\relax
\c*=#1\l*\c*\c*=.01\Lengthunit\m*\c*\divide\l*\m*
\c*=\the\Nhalfperiods5sp\multiply\c*\l*\N*\c*\divide\N*\*ths
\divide\N*2\multiply\N*2\advance\N*\*one
\dd*=.002\Lengthunit\dd*=#4\dd*\multiply\dd*\l*\divide\dd*\N*
\divide\dd*\*ths \multiply\dd*\magnitude
\d**=#1\multiply\N*4\divide\d**\N*\sm*\n*\*one\loop
\shl**{\dd*}\dt*=1.3\dd*\advance\n*\*one
\shl**{\dt*}\advance\n*\*one
\shl**{\dd*}\advance\n*\*two
\dd*-\dd*\ifnum\n*<\N*\repeat\n*\N*\shl**{0pt}}}

\def\wavebox#1{\setbox0\hbox{#1}\relax
\a*=\wd0\advance\a*14pt\b*=\ht0\advance\b*\dp0\advance\b*14pt\relax
\hbox{\kern9pt\relax
\rmov*(0pt,\ht0){\rmov*(-7pt,7pt){\wlin*\a*(1,0)[+]\wlin*\b*(0,-1)[-]}}\relax
\rmov*(\wd0,-\dp0){\rmov*(7pt,-7pt){\wlin*\a*(-1,0)[+]\wlin*\b*(0,1)[-]}}\relax
\box0\kern9pt}}

\def\rectangle#1(#2,#3){\relax
\lin#1(#2,0)\lin#1(0,#3)\mov#1(0,#3){\lin#1(#2,0)}\mov#1(#2,0){\lin#1(0,#3)}}

\def\dashrectangle#1(#2,#3){\dashlin#1(#2,0)\dashlin#1(0,#3)\relax
\mov#1(0,#3){\dashlin#1(#2,0)}\mov#1(#2,0){\dashlin#1(0,#3)}}

\def\waverectangle#1(#2,#3){\L*=#1\Lengthunit\a*=#2\L*\b*=#3\L*
\ifdim\a*<0pt\a*-\a*\def\x*{-1}\else\def\x*{1}\fi
\ifdim\b*<0pt\b*-\b*\def\y*{-1}\else\def\y*{1}\fi
\wlin*\a*(\x*,0)[-]\wlin*\b*(0,\y*)[+]\relax
\mov#1(0,#3){\wlin*\a*(\x*,0)[+]}\mov#1(#2,0){\wlin*\b*(0,\y*)[-]}}

\def\calcparab*{\ifnum\n*>\m*\k*\N*\advance\k*-\n*\else\k*\n*\fi
\a*=\the\k* sp\a*=10\a*\b*\dm*\advance\b*-\a*\k*\b*
\a*=\the\*ths\b*\divide\a*\l*\multiply\a*\k*
\divide\a*\l*\k*\*ths\r*\a*\advance\k*-\r*\dt*=\the\k*\L*}

\def\arcto#1(#2,#3)[#4]{\rlap{\toks0={#2}\toks1={#3}\calcnum*#1(#2,#3)\relax
\dm*=135sp\dm*=#1\dm*\d**=#1\Lengthunit\ifdim\dm*<0pt\dm*-\dm*\fi
\multiply\dm*\r*\a*=.3\dm*\a*=#4\a*\ifdim\a*<0pt\a*-\a*\fi
\advance\dm*\a*\N*\dm*\divide\N*10000\relax
\divide\N*2\multiply\N*2\advance\N*\*one
\L*=-.25\d**\L*=#4\L*\divide\d**\N*\divide\L*\*ths
\m*\N*\divide\m*2\dm*=\the\m*5sp\l*\dm*\sm*\n*\*one\loop
\calcparab*\shl**{-\dt*}\advance\n*1\ifnum\n*<\N*\repeat}}

\def\arrarcto#1(#2,#3)[#4]{\L*=#1\Lengthunit\L*=.54\L*
\arcto#1(#2,#3)[#4]\rmov*(#2\L*,#3\L*){\d*=.457\L*\d*=#4\d*\d**-\d*
\rmov*(#3\d**,#2\d*){\arrow.02(#2,#3)}}}

\def\dasharcto#1(#2,#3)[#4]{\rlap{\toks0={#2}\toks1={#3}\relax
\calcnum*#1(#2,#3)\dm*=\the\N*5sp\a*=.3\dm*\a*=#4\a*\ifdim\a*<0pt\a*-\a*\fi
\advance\dm*\a*\N*\dm*
\divide\N*20\multiply\N*2\advance\N*1\d**=#1\Lengthunit
\L*=-.25\d**\L*=#4\L*\divide\d**\N*\divide\L*\*ths
\m*\N*\divide\m*2\dm*=\the\m*5sp\l*\dm*
\sm*\n*\*one\loop\calcparab*
\shl**{-\dt*}\advance\n*1\ifnum\n*>\N*\else\calcparab*
\sh*(#2,#3){\xL*=#3\dt* \yL*=#2\dt*
\rx* \the\cos*\xL* \tmp* \the\sin*\yL* \advance\rx*\tmp*
\ry* \the\cos*\yL* \tmp* \the\sin*\xL* \advance\ry*-\tmp*
\kern\rx*\lower\ry*\hbox{\sm*}}\fi
\advance\n*1\ifnum\n*<\N*\repeat}}

\def\*shl*#1{\c*=\the\n*\d**\advance\c*#1\a**\d*\dt*\advance\d*#1\b**
\a*=\the\toks0\c*\b*=\the\toks1\d*\advance\a*-\b*
\b*=\the\toks1\c*\d*=\the\toks0\d*\advance\b*\d*
\rx* \the\cos*\a* \tmp* \the\sin*\b* \advance\rx*-\tmp*
\ry* \the\cos*\b* \tmp* \the\sin*\a* \advance\ry*\tmp*
\raise\ry*\rlap{\kern\rx*\unhcopy\spl*}}

\def\calcnormal*#1{\b**=10000sp\a**\b**\k*\n*\advance\k*-\m*
\multiply\a**\k*\divide\a**\m*\a**=#1\a**\ifdim\a**<0pt\a**-\a**\fi
\ifdim\a**>\b**\d*=.96\a**\advance\d*.4\b**
\else\d*=.96\b**\advance\d*.4\a**\fi
\d*=.01\d*\r*\d*\divide\a**\r*\divide\b**\r*
\ifnum\k*<0\a**-\a**\fi\d*=#1\d*\ifdim\d*<0pt\b**-\b**\fi
\k*\a**\a**=\the\k*\dd*\k*\b**\b**=\the\k*\dd*}

\def\wavearcto#1(#2,#3)[#4]{\rlap{\toks0={#2}\toks1={#3}\relax
\calcnum*#1(#2,#3)\c*=\the\N*5sp\a*=.4\c*\a*=#4\a*\ifdim\a*<0pt\a*-\a*\fi
\advance\c*\a*\N*\c*\divide\N*20\multiply\N*2\advance\N*-1\multiply\N*4\relax
\d**=#1\Lengthunit\dd*=.012\d**
\divide\dd*\*ths \multiply\dd*\magnitude
\ifdim\d**<0pt\d**-\d**\fi\L*=.25\d**
\divide\d**\N*\divide\dd*\N*\L*=#4\L*\divide\L*\*ths
\m*\N*\divide\m*2\dm*=\the\m*0sp\l*\dm*
\sm*\n*\*one\loop\calcnormal*{#4}\calcparab*
\*shl*{1}\advance\n*\*one\calcparab*
\*shl*{1.3}\advance\n*\*one\calcparab*
\*shl*{1}\advance\n*2\dd*-\dd*\ifnum\n*<\N*\repeat\n*\N*\shl**{0pt}}}

\def\triangarcto#1(#2,#3)[#4]{\rlap{\toks0={#2}\toks1={#3}\relax
\calcnum*#1(#2,#3)\c*=\the\N*5sp\a*=.4\c*\a*=#4\a*\ifdim\a*<0pt\a*-\a*\fi
\advance\c*\a*\N*\c*\divide\N*20\multiply\N*2\advance\N*-1\multiply\N*2\relax
\d**=#1\Lengthunit\dd*=.012\d**
\divide\dd*\*ths \multiply\dd*\magnitude
\ifdim\d**<0pt\d**-\d**\fi\L*=.25\d**
\divide\d**\N*\divide\dd*\N*\L*=#4\L*\divide\L*\*ths
\m*\N*\divide\m*2\dm*=\the\m*0sp\l*\dm*
\sm*\n*\*one\loop\calcnormal*{#4}\calcparab*
\*shl*{1}\advance\n*2\dd*-\dd*\ifnum\n*<\N*\repeat\n*\N*\shl**{0pt}}}

\def\hr*#1{\L*=\xscale\Lengthunit\ifnum
\angle**=0\clap{\vrule width#1\L* height.1pt}\else
\L*=#1\L*\L*=.5\L*\rmov*(-\L*,0pt){\sm*}\rmov*(\L*,0pt){\sl*}\fi}

\def\shade#1[#2]{\rlap{\Lengthunit=#1\Lengthunit
\special{em:linewidth .001pt}\relax
\mov(0,#2.05){\hr*{.994}}\mov(0,#2.1){\hr*{.980}}\relax
\mov(0,#2.15){\hr*{.953}}\mov(0,#2.2){\hr*{.916}}\relax
\mov(0,#2.25){\hr*{.867}}\mov(0,#2.3){\hr*{.798}}\relax
\mov(0,#2.35){\hr*{.715}}\mov(0,#2.4){\hr*{.603}}\relax
\mov(0,#2.45){\hr*{.435}}\special{em:linewidth \the\linwid*}}}

\def\dshade#1[#2]{\rlap{\special{em:linewidth .001pt}\relax
\Lengthunit=#1\Lengthunit\if#2-\def\t*{+}\else\def\t*{-}\fi
\mov(0,\t*.025){\relax
\mov(0,#2.05){\hr*{.995}}\mov(0,#2.1){\hr*{.988}}\relax
\mov(0,#2.15){\hr*{.969}}\mov(0,#2.2){\hr*{.937}}\relax
\mov(0,#2.25){\hr*{.893}}\mov(0,#2.3){\hr*{.836}}\relax
\mov(0,#2.35){\hr*{.760}}\mov(0,#2.4){\hr*{.662}}\relax
\mov(0,#2.45){\hr*{.531}}\mov(0,#2.5){\hr*{.320}}\relax
\special{em:linewidth \the\linwid*}}}}

\def\vdot{\rlap{\kern-1.9pt\lower1.8pt\hbox{$\scriptstyle\bullet$}}}
\def\vtimes{\rlap{\kern-3pt\lower1.8pt\hbox{$\scriptstyle\times$}}}
\def\vDot{\rlap{\kern-2.3pt\lower2.7pt\hbox{$\bullet$}}}
\def\vTimes{\rlap{\kern-3.6pt\lower2.4pt\hbox{$\times$}}}

\def\arc(#1)[#2,#3]{{\k*=#2\l*=#3\m*=\l*
\advance\m*-6\ifnum\k*>\l*\relax\else
{\rotate(#2)\mov(#1,0){\sm*}}\loop
\ifnum\k*<\m*\advance\k*5{\rotate(\k*)\mov(#1,0){\sl*}}\repeat
{\rotate(#3)\mov(#1,0){\sl*}}\fi}}

\def\dasharc(#1)[#2,#3]{{\k**=#2\n*=#3\advance\n*-1\advance\n*-\k**
\L*=1000sp\L*#1\L* \multiply\L*\n* \multiply\L*\Nhalfperiods
\divide\L*57\N*\L* \divide\N*2000\ifnum\N*=0\N*1\fi
\r*\n*  \divide\r*\N* \ifnum\r*<2\r*2\fi
\m**\r* \divide\m**2 \l**\r* \advance\l**-\m** \N*\n* \divide\N*\r*
\k**\r* \multiply\k**\N* \dn*\n* \advance\dn*-\k** \divide\dn*2\advance\dn*\*one
\r*\l** \divide\r*2\advance\dn*\r* \advance\N*-2\k**#2\relax
\ifnum\l**<6{\rotate(#2)\mov(#1,0){\sm*}}\advance\k**\dn*
{\rotate(\k**)\mov(#1,0){\sl*}}\advance\k**\m**
{\rotate(\k**)\mov(#1,0){\sm*}}\loop
\advance\k**\l**{\rotate(\k**)\mov(#1,0){\sl*}}\advance\k**\m**
{\rotate(\k**)\mov(#1,0){\sm*}}\advance\N*-1\ifnum\N*>0\repeat
{\rotate(#3)\mov(#1,0){\sl*}}\else\advance\k**\dn*
\arc(#1)[#2,\k**]\loop\advance\k**\m** \r*\k**
\advance\k**\l** {\arc(#1)[\r*,\k**]}\relax
\advance\N*-1\ifnum\N*>0\repeat
\advance\k**\m**\arc(#1)[\k**,#3]\fi}}

\def\triangarc#1(#2)[#3,#4]{{\k**=#3\n*=#4\advance\n*-\k**
\L*=1000sp\L*#2\L* \multiply\L*\n* \multiply\L*\Nhalfperiods
\divide\L*57\N*\L* \divide\N*1000\ifnum\N*=0\N*1\fi
\d**=#2\Lengthunit \d*\d** \divide\d*57\multiply\d*\n*
\r*\n*  \divide\r*\N* \ifnum\r*<2\r*2\fi
\m**\r* \divide\m**2 \l**\r* \advance\l**-\m** \N*\n* \divide\N*\r*
\dt*\d* \divide\dt*\N* \dt*.5\dt* \dt*#1\dt*
\divide\dt*1000\multiply\dt*\magnitude
\k**\r* \multiply\k**\N* \dn*\n* \advance\dn*-\k** \divide\dn*2\relax
\r*\l** \divide\r*2\advance\dn*\r* \advance\N*-1\k**#3\relax
{\rotate(#3)\mov(#2,0){\sm*}}\advance\k**\dn*
{\rotate(\k**)\mov(#2,0){\sl*}}\advance\k**-\m**\advance\l**\m**\loop\dt*-\dt*
\d*\d** \advance\d*\dt*
\advance\k**\l**{\rotate(\k**)\rmov*(\d*,0pt){\sl*}}%
\advance\N*-1\ifnum\N*>0\repeat\advance\k**\m**
{\rotate(\k**)\mov(#2,0){\sl*}}{\rotate(#4)\mov(#2,0){\sl*}}}}

\def\wavearc#1(#2)[#3,#4]{{\k**=#3\n*=#4\advance\n*-\k**
\L*=4000sp\L*#2\L* \multiply\L*\n* \multiply\L*\Nhalfperiods
\divide\L*57\N*\L* \divide\N*1000\ifnum\N*=0\N*1\fi
\d**=#2\Lengthunit \d*\d** \divide\d*57\multiply\d*\n*
\r*\n*  \divide\r*\N* \ifnum\r*=0\r*1\fi
\m**\r* \divide\m**2 \l**\r* \advance\l**-\m** \N*\n* \divide\N*\r*
\dt*\d* \divide\dt*\N* \dt*.7\dt* \dt*#1\dt*
\divide\dt*1000\multiply\dt*\magnitude
\k**\r* \multiply\k**\N* \dn*\n* \advance\dn*-\k** \divide\dn*2\relax
\divide\N*4\advance\N*-1\k**#3\relax
{\rotate(#3)\mov(#2,0){\sm*}}\advance\k**\dn*
{\rotate(\k**)\mov(#2,0){\sl*}}\advance\k**-\m**\advance\l**\m**\loop\dt*-\dt*
\d*\d** \advance\d*\dt* \dd*\d** \advance\dd*1.3\dt*
\advance\k**\r*{\rotate(\k**)\rmov*(\d*,0pt){\sl*}}\relax
\advance\k**\r*{\rotate(\k**)\rmov*(\dd*,0pt){\sl*}}\relax
\advance\k**\r*{\rotate(\k**)\rmov*(\d*,0pt){\sl*}}\relax
\advance\k**\r*
\advance\N*-1\ifnum\N*>0\repeat\advance\k**\m**
{\rotate(\k**)\mov(#2,0){\sl*}}{\rotate(#4)\mov(#2,0){\sl*}}}}

\def\gmov*#1(#2,#3)#4{\rlap{\L*=#1\Lengthunit
\xL*=#2\L* \yL*=#3\L*
\rx* \gcos*\xL* \tmp* \gsin*\yL* \advance\rx*-\tmp*
\ry* \gcos*\yL* \tmp* \gsin*\xL* \advance\ry*\tmp*
\rx*=\xscale\rx* \ry*=\yscale\ry*
\xL* \the\cos*\rx* \tmp* \the\sin*\ry* \advance\xL*-\tmp*
\yL* \the\cos*\ry* \tmp* \the\sin*\rx* \advance\yL*\tmp*
\kern\xL*\raise\yL*\hbox{#4}}}

\def\rgmov*(#1,#2)#3{\rlap{\xL*#1\yL*#2\relax
\rx* \gcos*\xL* \tmp* \gsin*\yL* \advance\rx*-\tmp*
\ry* \gcos*\yL* \tmp* \gsin*\xL* \advance\ry*\tmp*
\rx*=\xscale\rx* \ry*=\yscale\ry*
\xL* \the\cos*\rx* \tmp* \the\sin*\ry* \advance\xL*-\tmp*
\yL* \the\cos*\ry* \tmp* \the\sin*\rx* \advance\yL*\tmp*
\kern\xL*\raise\yL*\hbox{#3}}}

\def\Earc(#1)[#2,#3][#4,#5]{{\k*=#2\l*=#3\m*=\l*
\advance\m*-6\ifnum\k*>\l*\relax\else\def\xscale{#4}\def\yscale{#5}\relax
{\angle**0\rotate(#2)}\gmov*(#1,0){\sm*}\loop
\ifnum\k*<\m*\advance\k*5\relax
{\angle**0\rotate(\k*)}\gmov*(#1,0){\sl*}\repeat
{\angle**0\rotate(#3)}\gmov*(#1,0){\sl*}\relax
\def\xscale{1}\def\yscale{1}\fi}}

\def\dashEarc(#1)[#2,#3][#4,#5]{{\k**=#2\n*=#3\advance\n*-1\advance\n*-\k**
\L*=1000sp\L*#1\L* \multiply\L*\n* \multiply\L*\Nhalfperiods
\divide\L*57\N*\L* \divide\N*2000\ifnum\N*=0\N*1\fi
\r*\n*  \divide\r*\N* \ifnum\r*<2\r*2\fi
\m**\r* \divide\m**2 \l**\r* \advance\l**-\m** \N*\n* \divide\N*\r*
\k**\r*\multiply\k**\N* \dn*\n* \advance\dn*-\k** \divide\dn*2\advance\dn*\*one
\r*\l** \divide\r*2\advance\dn*\r* \advance\N*-2\k**#2\relax
\ifnum\l**<6\def\xscale{#4}\def\yscale{#5}\relax
{\angle**0\rotate(#2)}\gmov*(#1,0){\sm*}\advance\k**\dn*
{\angle**0\rotate(\k**)}\gmov*(#1,0){\sl*}\advance\k**\m**
{\angle**0\rotate(\k**)}\gmov*(#1,0){\sm*}\loop
\advance\k**\l**{\angle**0\rotate(\k**)}\gmov*(#1,0){\sl*}\advance\k**\m**
{\angle**0\rotate(\k**)}\gmov*(#1,0){\sm*}\advance\N*-1\ifnum\N*>0\repeat
{\angle**0\rotate(#3)}\gmov*(#1,0){\sl*}\def\xscale{1}\def\yscale{1}\else
\advance\k**\dn* \Earc(#1)[#2,\k**][#4,#5]\loop\advance\k**\m** \r*\k**
\advance\k**\l** {\Earc(#1)[\r*,\k**][#4,#5]}\relax
\advance\N*-1\ifnum\N*>0\repeat
\advance\k**\m**\Earc(#1)[\k**,#3][#4,#5]\fi}}

\def\triangEarc#1(#2)[#3,#4][#5,#6]{{\k**=#3\n*=#4\advance\n*-\k**
\L*=1000sp\L*#2\L* \multiply\L*\n* \multiply\L*\Nhalfperiods
\divide\L*57\N*\L* \divide\N*1000\ifnum\N*=0\N*1\fi
\d**=#2\Lengthunit \d*\d** \divide\d*57\multiply\d*\n*
\r*\n*  \divide\r*\N* \ifnum\r*<2\r*2\fi
\m**\r* \divide\m**2 \l**\r* \advance\l**-\m** \N*\n* \divide\N*\r*
\dt*\d* \divide\dt*\N* \dt*.5\dt* \dt*#1\dt*
\divide\dt*1000\multiply\dt*\magnitude
\k**\r* \multiply\k**\N* \dn*\n* \advance\dn*-\k** \divide\dn*2\relax
\r*\l** \divide\r*2\advance\dn*\r* \advance\N*-1\k**#3\relax
\def\xscale{#5}\def\yscale{#6}\relax
{\angle**0\rotate(#3)}\gmov*(#2,0){\sm*}\advance\k**\dn*
{\angle**0\rotate(\k**)}\gmov*(#2,0){\sl*}\advance\k**-\m**
\advance\l**\m**\loop\dt*-\dt* \d*\d** \advance\d*\dt*
\advance\k**\l**{\angle**0\rotate(\k**)}\rgmov*(\d*,0pt){\sl*}\relax
\advance\N*-1\ifnum\N*>0\repeat\advance\k**\m**
{\angle**0\rotate(\k**)}\gmov*(#2,0){\sl*}\relax
{\angle**0\rotate(#4)}\gmov*(#2,0){\sl*}\def\xscale{1}\def\yscale{1}}}

\def\waveEarc#1(#2)[#3,#4][#5,#6]{{\k**=#3\n*=#4\advance\n*-\k**
\L*=4000sp\L*#2\L* \multiply\L*\n* \multiply\L*\Nhalfperiods
\divide\L*57\N*\L* \divide\N*1000\ifnum\N*=0\N*1\fi
\d**=#2\Lengthunit \d*\d** \divide\d*57\multiply\d*\n*
\r*\n*  \divide\r*\N* \ifnum\r*=0\r*1\fi
\m**\r* \divide\m**2 \l**\r* \advance\l**-\m** \N*\n* \divide\N*\r*
\dt*\d* \divide\dt*\N* \dt*.7\dt* \dt*#1\dt*
\divide\dt*1000\multiply\dt*\magnitude
\k**\r* \multiply\k**\N* \dn*\n* \advance\dn*-\k** \divide\dn*2\relax
\divide\N*4\advance\N*-1\k**#3\def\xscale{#5}\def\yscale{#6}\relax
{\angle**0\rotate(#3)}\gmov*(#2,0){\sm*}\advance\k**\dn*
{\angle**0\rotate(\k**)}\gmov*(#2,0){\sl*}\advance\k**-\m**
\advance\l**\m**\loop\dt*-\dt*
\d*\d** \advance\d*\dt* \dd*\d** \advance\dd*1.3\dt*
\advance\k**\r*{\angle**0\rotate(\k**)}\rgmov*(\d*,0pt){\sl*}\relax
\advance\k**\r*{\angle**0\rotate(\k**)}\rgmov*(\dd*,0pt){\sl*}\relax
\advance\k**\r*{\angle**0\rotate(\k**)}\rgmov*(\d*,0pt){\sl*}\relax
\advance\k**\r*
\advance\N*-1\ifnum\N*>0\repeat\advance\k**\m**
{\angle**0\rotate(\k**)}\gmov*(#2,0){\sl*}\relax
{\angle**0\rotate(#4)}\gmov*(#2,0){\sl*}\def\xscale{1}\def\yscale{1}}}

\newcount\CatcodeOfAtSign
\CatcodeOfAtSign=\the\catcode`\@
\catcode`\@=11
\def\@arc#1[#2][#3]{\rlap{\Lengthunit=#1\Lengthunit
\sm*\l*arc(#2.1914,#3.0381)[#2][#3]\relax
\mov(#2.1914,#3.0381){\l*arc(#2.1622,#3.1084)[#2][#3]}\relax
\mov(#2.3536,#3.1465){\l*arc(#2.1084,#3.1622)[#2][#3]}\relax
\mov(#2.4619,#3.3086){\l*arc(#2.0381,#3.1914)[#2][#3]}}}

\def\dash@arc#1[#2][#3]{\rlap{\Lengthunit=#1\Lengthunit
\d*arc(#2.1914,#3.0381)[#2][#3]\relax
\mov(#2.1914,#3.0381){\d*arc(#2.1622,#3.1084)[#2][#3]}\relax
\mov(#2.3536,#3.1465){\d*arc(#2.1084,#3.1622)[#2][#3]}\relax
\mov(#2.4619,#3.3086){\d*arc(#2.0381,#3.1914)[#2][#3]}}}

\def\wave@arc#1[#2][#3]{\rlap{\Lengthunit=#1\Lengthunit
\w*lin(#2.1914,#3.0381)\relax
\mov(#2.1914,#3.0381){\w*lin(#2.1622,#3.1084)}\relax
\mov(#2.3536,#3.1465){\w*lin(#2.1084,#3.1622)}\relax
\mov(#2.4619,#3.3086){\w*lin(#2.0381,#3.1914)}}}

\def\bezier#1(#2,#3)(#4,#5)(#6,#7){\N*#1\l*\N* \advance\l*\*one
\d* #4\Lengthunit \advance\d* -#2\Lengthunit \multiply\d* \*two
\b* #6\Lengthunit \advance\b* -#2\Lengthunit
\advance\b*-\d* \divide\b*\N*
\d** #5\Lengthunit \advance\d** -#3\Lengthunit \multiply\d** \*two
\b** #7\Lengthunit \advance\b** -#3\Lengthunit
\advance\b** -\d** \divide\b**\N*
\mov(#2,#3){\sm*{\loop\ifnum\m*<\l*
\a*\m*\b* \advance\a*\d* \divide\a*\N* \multiply\a*\m*
\a**\m*\b** \advance\a**\d** \divide\a**\N* \multiply\a**\m*
\rmov*(\a*,\a**){\unhcopy\spl*}\advance\m*\*one\repeat}}}

\catcode`\*=12

\newcount\n@ast
\def\n@ast@#1{\n@ast0\relax\get@ast@#1\end}
\def\get@ast@#1{\ifx#1\end\let\next\relax\else
\ifx#1*\advance\n@ast1\fi\let\next\get@ast@\fi\next}

\newif\if@up \newif\if@dwn
\def\up@down@#1{\@upfalse\@dwnfalse
\if#1u\@uptrue\fi\if#1U\@uptrue\fi\if#1+\@uptrue\fi
\if#1d\@dwntrue\fi\if#1D\@dwntrue\fi\if#1-\@dwntrue\fi}

\def\halfcirc#1(#2)[#3]{{\Lengthunit=#2\Lengthunit\up@down@{#3}\relax
\if@up\mov(0,.5){\@arc[-][-]\@arc[+][-]}\fi
\if@dwn\mov(0,-.5){\@arc[-][+]\@arc[+][+]}\fi
\def\lft{\mov(0,.5){\@arc[-][-]}\mov(0,-.5){\@arc[-][+]}}\relax
\def\rght{\mov(0,.5){\@arc[+][-]}\mov(0,-.5){\@arc[+][+]}}\relax
\if#3l\lft\fi\if#3L\lft\fi\if#3r\rght\fi\if#3R\rght\fi
\n@ast@{#1}\relax
\ifnum\n@ast>0\if@up\shade[+]\fi\if@dwn\shade[-]\fi\fi
\ifnum\n@ast>1\if@up\dshade[+]\fi\if@dwn\dshade[-]\fi\fi}}

\def\halfdashcirc(#1)[#2]{{\Lengthunit=#1\Lengthunit\up@down@{#2}\relax
\if@up\mov(0,.5){\dash@arc[-][-]\dash@arc[+][-]}\fi
\if@dwn\mov(0,-.5){\dash@arc[-][+]\dash@arc[+][+]}\fi
\def\lft{\mov(0,.5){\dash@arc[-][-]}\mov(0,-.5){\dash@arc[-][+]}}\relax
\def\rght{\mov(0,.5){\dash@arc[+][-]}\mov(0,-.5){\dash@arc[+][+]}}\relax
\if#2l\lft\fi\if#2L\lft\fi\if#2r\rght\fi\if#2R\rght\fi}}

\def\halfwavecirc(#1)[#2]{{\Lengthunit=#1\Lengthunit\up@down@{#2}\relax
\if@up\mov(0,.5){\wave@arc[-][-]\wave@arc[+][-]}\fi
\if@dwn\mov(0,-.5){\wave@arc[-][+]\wave@arc[+][+]}\fi
\def\lft{\mov(0,.5){\wave@arc[-][-]}\mov(0,-.5){\wave@arc[-][+]}}\relax
\def\rght{\mov(0,.5){\wave@arc[+][-]}\mov(0,-.5){\wave@arc[+][+]}}\relax
\if#2l\lft\fi\if#2L\lft\fi\if#2r\rght\fi\if#2R\rght\fi}}

\catcode`\*=11

\def\Circle#1(#2){\halfcirc#1(#2)[u]\halfcirc#1(#2)[d]\n@ast@{#1}\relax
\ifnum\n@ast>0\L*=\xscale\Lengthunit
\ifnum\angle**=0\clap{\vrule width#2\L* height.1pt}\else
\L*=#2\L*\L*=.5\L*\special{em:linewidth .001pt}\relax
\rmov*(-\L*,0pt){\sm*}\rmov*(\L*,0pt){\sl*}\relax
\special{em:linewidth \the\linwid*}\fi\fi}

\catcode`\*=12

\def\wavecirc(#1){\halfwavecirc(#1)[u]\halfwavecirc(#1)[d]}

\def\dashcirc(#1){\halfdashcirc(#1)[u]\halfdashcirc(#1)[d]}

\def\xscale{1}
\def\yscale{1}

\def\Ellipse#1(#2)[#3,#4]{\def\xscale{#3}\def\yscale{#4}\relax
\Circle#1(#2)\def\xscale{1}\def\yscale{1}}

\def\dashEllipse(#1)[#2,#3]{\def\xscale{#2}\def\yscale{#3}\relax
\dashcirc(#1)\def\xscale{1}\def\yscale{1}}

\def\waveEllipse(#1)[#2,#3]{\def\xscale{#2}\def\yscale{#3}\relax
\wavecirc(#1)\def\xscale{1}\def\yscale{1}}

\def\halfEllipse#1(#2)[#3][#4,#5]{\def\xscale{#4}\def\yscale{#5}\relax
\halfcirc#1(#2)[#3]\def\xscale{1}\def\yscale{1}}

\def\halfdashEllipse(#1)[#2][#3,#4]{\def\xscale{#3}\def\yscale{#4}\relax
\halfdashcirc(#1)[#2]\def\xscale{1}\def\yscale{1}}

\def\halfwaveEllipse(#1)[#2][#3,#4]{\def\xscale{#3}\def\yscale{#4}\relax
\halfwavecirc(#1)[#2]\def\xscale{1}\def\yscale{1}}

\catcode`\@=\the\CatcodeOfAtSign

\section{Introduction}

The precise investigation of the energy levels of hydrogenic atoms
(muonium, positronium, hydrogen atom, deuterium, helium ions et al.)
allows to obtain more exact values for many fundamental physical constants
such as the lepton masses, the ratio of lepton and proton masses,
the fine structure constant, the Rydberg constant which are used for
creating standards of units \cite{MT,SGK}. The insertion of new simple
atomic systems in the range of experimental investigation can lead to
significant progress in solving of these problems. Muonic hydrogen $(\mu p)$
is just one of a number of simple atoms which has attracted considerable
attention in the last years. Since the muon is heavier than the electron
by a factor of 206, the energy structure of $(\mu p)$ is extremely sensitive
to the effect of the electron vacuum polarization, recoil effect, the proton
structure and polarizability corrections \cite{EGS,BR,P2,Borie}. In the Lamb
shift $(2P-2S)$ of $(\mu p)$ the value of the proton structure correction
of order $(Z\alpha)^4$ increases essentially as compared with electronic
hydrogen. Considering that this contribution is determined also by the proton
charge radius $r_p$, the experimental investigation of the energy spectrum
of muonic hydrogen can play the key role in a more precise study of the proton
structure along with the experimental data on the electron-nucleon scattering
\cite{SGK,Simon,Sick}. At present time the Lamb shift $(2P-2S)$ experiment
in muonic hydrogen at PSI (Paul Sherrer Institute)
with a precision of 30 ppm entered the closing stage \cite{FK1,FK2}.
The suggested aim is to determine the root-mean-square (rms) charge radius
of the proton to the accuracy $10^{-3}$, about a factor 20 times better than presently
known from electron scattering experiments. It opens the possibility to
check bound state QED predictions toward a level of $10^{-7}$ precision.
Because the Lamb shift experiment in $(\mu p)$ measure the energy difference
between the $2^3S_{1/2}$ and $2^5P_{3/2}$ atomic levels, it is important
to know the hyperfine structure of $2S$ and $2P$ energy levels \cite{P2,apm2005}.
In this connection it is important to consider again several theoretical
contributions which have essential role in the calculation
of the total theoretical value of the $(2P-2S)$ Lamb shift in the atom
$(\mu p)$ with the necessary accuracy. The proton polarizability
contribution is among important effects \cite{DS,BE,SPH,IBK,RR,FM1,P1},
which was calculated earlier in muonic hydrogen on the basis of experimental
data on the structure functions of the lepton-nucleon scattering.
Slightly different values for this contribution occur to the present
\cite{SPH,RR,FM1,P1}.

The study of electromagnetic excitations of baryonic resonances
which is carried out at CEBAF (Continuous Electron Beam
Accelerator Facility), entered currently a new phase of
development because new data with unprecedented precision have
become available. The aims of the CEBAF investigations are in the
measurements of the nucleon transition form factors to nucleon
resonances $N^\ast$ at different photon virtualities $Q^2$ in the
resonance region, in the study of the gluon content of baryonic
resonances and the helicity amplitudes $A_{1/2}$, $A_{3/2}$,
$S_{1/2}$ for different states $N^\ast$ \cite{Burkert}. Recently,
precise measurements for exclusive electroproduction of $\pi^0$,
$\pi^+$, $\eta$ mesons on protons in the resonance region were
obtained at Jefferson Lab using the CEBAF Large Acceptance
Spectrometer (CLAS) \cite{CLAS}. The obtained experimental data
include measurements of the cross sections, angular distributions
for the $n\pi^+$, $p\eta$ final states. These data allow also to
investigate the contributions of resonances to $\pi$, $\eta$
electroproduction in detail. The performed experiments permitted
already to refine the values of a number of theoretical parameters
determining the production of the nucleon resonances $N^\ast$ in
the reaction $\gamma^\ast+N\to N^\ast$. The goal of the present
investigation is to perform new calculation of the proton
polarizability contribution to the Lamb shift in hydrogen atom
based on theoretical model (unitary isobar model), describing the
processes of photo- and electroproduction of $\pi,\eta$ mesons,
nucleon resonances on the nucleon in the resonance region, and on
evolution equations for the parton distributions in deep inelastic
region.

\section{General formalism}

The proton polarizability contribution to the Lamb shift of order
$(Z\alpha)^5$ is determined by the amplitude of virtual Compton
forward scattering $\gamma^\ast+p\to\gamma^\ast+p$, presented on the
Feynman diagrams in Fig.1. It's parameterization has the following form
\cite{RF,JB}:
\begin{equation}
M_{\mu\nu}^{(p)}=\bar v(p_2)\Biggl\{\frac{1}{2}C_1\left(-g_{\mu\nu}+\frac{k_\mu k_\nu}
{k^2}\right)+\frac{1}{2m^2_2}C_2\left(p_{2 \mu}-\frac{m_2\nu}{k^2}k_\mu\right)
\left(p_{2 \nu}-\frac{m_2\nu}{k^2}k_\nu\right)+
\end{equation}
\begin{displaymath}
+\frac{1}{2m_2}H_1\left([\gamma_\nu,\hat k]p_{2
\mu}-[\gamma_\mu,\hat k]p_{2 \nu}+[\gamma_\mu, \gamma_\nu]\right)
+\frac{1}{2}H_2\left([\gamma_\nu,\hat k]k_\mu-[\gamma_\mu,\hat
k]k_\nu+ [\gamma_\mu,\gamma_\nu]k^2\right)\Biggr\}v(q_2),
\end{displaymath}
where $k$ is the four-momentum of the virtual photon, $\nu=k_0$ is the virtual
photon energy, $m_2$ is the proton mass. Symmetrical part of the tensor (1)
gives the contribution to the Lamb shift (the structure functions
$C_{1,2}(\nu, k^2)$) and antisymmetric part contributes to the hyperfine
structure (the structure functions $H_{1,2}(\nu,k^2)$). The lepton tensor
is the following:
\begin{equation}
M_{\mu\nu}^{(l)}=\bar u(q_1)\left[\gamma_\mu\frac{\hat{p_1}+\hat k+m_1}
{(p_1+k)^2-m_1^2+i\epsilon}\gamma_\nu+\gamma_\nu\frac{\hat{p_1}-\hat k+m_1}
{(p_1-k)^2-m_1^2+i\epsilon}\gamma_\mu\right]u(p_1),
\end{equation}
where $p_{1,2}$, $q_{1,2}$ are four-momenta of the lepton and proton
in the initial and final states, $m_1$ is the lepton mass.

\begin{figure*}[t!]
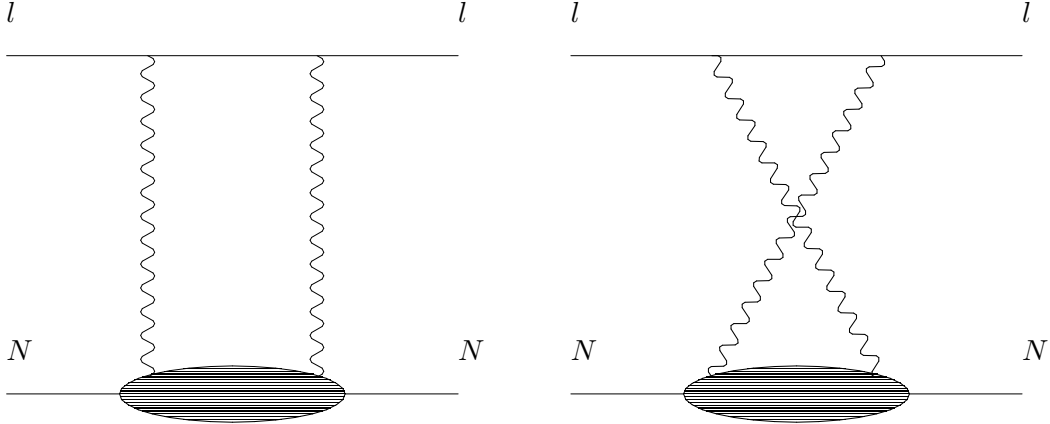

%\setcaptionmargin{5mm}
%\onelinecaptionsfalse
\magnitude=2000
\GRAPH(hsize=15){
\mov(0,0){\lin(1,0)}%
\mov(0,0.3){$N$}%
\mov(0.,3.3){$l$}%
\mov(4,0.3){$N$}%
\mov(4,3.3){$l$}%
\mov(5,0.3){$N$}%
\mov(5.,3.3){$l$}%
\mov(9,0.3){$N$}%
\mov(9,3.3){$l$}%
\mov(3,0){\lin(1,0)}%
\mov(5,0){\lin(1,0)}%
\mov(8,0){\lin(1,0)}%
\mov(0,3){\lin(4,0)}%
\mov(5,3){\lin(4,0)}%
\mov(1.25,3){\wavelin(0,-2.85)}%
%\mov(2.,3.){\halfwavecirc(0.7)[D]}%
\mov(2.75,3){\wavelin(0,-2.85)}%
\mov(2.,0){\Ellipse*(0.5)[4,1]}%
\mov(7.,0){\Ellipse*(0.5)[4,1]}%
\mov(7.75,3){\wavelin(-1.5,-2.85)}%
%\mov(7.,3.){\halfwavecirc(0.7)[D]}%
\mov(6.25,3){\wavelin(1.5,-2.85)}%
}
\bigskip
%\captionstyle{normal}
%\onelinecaptionsfalse
\caption{Two-photon Feynman amplitudes determining the correction on the
proton polarizability in the Lamb shift of hydrogen atom.}
\end{figure*}

Taking the product of the amplitudes (1) and (2), we can extract the part
of the interaction operator which is not dependent on the particle spins
and contributes to the Lamb shift (LS):
\begin{equation}
\left[M_{\mu\nu}^{(\mu)}M_{\mu\nu}^{(p)}\right]^{LS}=\frac{2m_1}
{k^4-4k_0^2m_1^2}\left[-(2k_0^2+k^2)C_1+(k^2-k_0^2)C_2\right].
\end{equation}
The structure functions $C_i(k_0, k^2)$ satisfy a following dispersion
relations \cite{DPV}:
\begin{equation}
C_1(k_0,k^2)=C_1(0,k^2)+\frac{1}{\pi}k_0^2\int_{\nu_0}^\infty\frac{d\nu^2}
{\nu^2(\nu^2-k_0^2)}Im C_1(\nu, k^2),
\end{equation}
\begin{equation}
C_2(k_0,k^2)=\frac{1}{\pi}\int_{\nu_0}^\infty\frac{d\nu^2}
{(\nu^2-k_0^2)}Im C_2(\nu, k^2),
\end{equation}
\begin{displaymath}
\nu_0=m_\pi+\frac{1}{2m_2}(Q^2+m_\pi^2),~~Q^2=-k^2.
\end{displaymath}
The threshold value of the photon energy $\nu_0$ represents the minimal
energy needed for the production of the $\pi$-meson in the reaction
$\gamma^\ast+p\to\pi^0+p$. Let us to point out that reliable data on the
subtraction term in the first dispersion integral (4) are absent. But in
the limit of small values of $k^2$ this term is connected with the proton
magnetic polarizability:
\begin{equation}
\lim_{k^2\rightarrow 0}\frac{C_1(0,k^2)}{k^2}=\frac{m_2}{\alpha}\beta_M,
\end{equation}
where $\beta_M=1.9(0.5)\times 10^{-4}~fm^3$ \cite{PDG}. The dipole
parameterization for $\beta_M(k^2)$ was suggested in Ref.\cite{P1}:
\begin{equation}
\beta_M(k^2)=\beta_M\frac{\Lambda^8}{(\Lambda^2+k^2)^4},
\end{equation}
where $\Lambda^2=0.71~GeV^2$ as for the elastic nucleon form
factor. Imaginary parts of the amplitudes $C_i(k_0,k^2)$ are
expressed in terms of the structure functions $F_i(x,Q^2)$ for
deep inelastic scattering as
\begin{equation}
\frac{1}{\pi}Im C_1(x,Q^2)=\frac{F_1(x,Q^2)}{m_2},~~~\frac{1}
{\pi}Im C_2(x,Q^2)=\frac{F_2(x,Q^2)}{\nu},~~x=\frac{Q^2}{2m_2\nu}.
\end{equation}
So, to obtain numerical value of the proton polarizability correction
we can use experimental data for the functions $F_{1,2}(\nu,k^2)$ and
different parameterizations for it prepared on the basis of these results
\cite{BK,Brasse,AL}.
Making use of relations (5)-(7) and transforming the integration in the loop
amplitudes to four-dimensional Euclidean space with the aid of the formula
\begin{equation}
\int d^4k=4\pi\int_0^\infty k^3dk\int_0^\pi\sin^2\phi\cdot d\phi,~~k^0=k\cos\phi,
\end{equation}
we can perform the integration over the angle variable $\phi$ and represent
the proton polarizability contribution to the Lamb shift of hydrogen atom
in the form:
\begin{equation}
\Delta E^{LS}_{pol}=-\frac{2\mu^3(Z\alpha)^5}{\pi n^3m_1^4}\int_0^\infty
d k\int_{\nu_0}^\infty dy ~{\cal F}(y,k)
+\frac{2\mu^3(Z\alpha)^4}{\pi n^3m_1}\int_0^\infty h(k^2)\beta_M(k^2)kdk,
\end{equation}
\begin{equation}
{\cal F}(y,k)=\frac{1}{(R+1)st^2(4s-t)}\Biggl\{-8\sqrt{s}(1+s)^{3/2}
\sqrt{t}(2s+R)-
\end{equation}
\begin{displaymath}
-\sqrt{t}(t-4s)[t+s(6+2R+4s+t)]+\sqrt{4+t}[(t-2)t+
s\left(8+t^2+2R(t+4)\right)]\Biggr\} F_2(y,k^2),
\end{displaymath}
\begin{equation}
h(k^2)=1+\left(1-\frac{t}{2}\right)\left(\sqrt{\frac{4}{t}+1}-1\right),~~t=
\frac{k^2}{m_1^2},~~s=\frac{y^2}{k^2},
\end{equation}
where $R(y,k^2)=\sigma_L/\sigma_T$ is the ratio of the cross sections
for the absorption of longitudinally and transversely polarized photons
by hadrons. In such a way, the correction $\Delta E^{LS}_{pol}$ is
expressed in terms of two structure functions $F_2(\nu,k^2)$ and
$R(\nu,k^2)$, describing unpolarized lepton-nucleon scattering.

\section{Structure functions of unpolarized lepton-nucleon scattering}

The greatest contribution to the integral (10) is given by the
region of the variable $k^2$: $0 \div  1~GeV^2$ and near the
threshold values of a photon energy $\nu$. So, the exact
construction of the structure functions $F_2$, $R$ in this region
is very important to obtain reliable estimation of the proton
polarizability effect. Deep inelastic lepton-nucleon scattering is
described by the following reaction:
\begin{equation}
l+N\to l'+X,
\end{equation}
where $X$ denotes the sum over all final particles. This reaction represents
the inclusive lepton $(l)$ production with the measurement of their energy
and scattering angle. It is assumed that this process occurs due to one-photon
exchange. An important kinematical variable of the reaction (13) is the
invariant mass of electroproduced hadronic system $W$:
\begin{equation}
W^2=m_2^2-Q^2+2m_2\nu,~~k^2=-Q^2.
\end{equation}
Using the variable (14) we can divide the total integration
region in Eq.(10) on the resonance region $W\leq 2$ GeV where the
production of low-lying nucleon resonances occurs and deep
inelastic region when $W > 2$ GeV.

\begin{figure*}[t!]
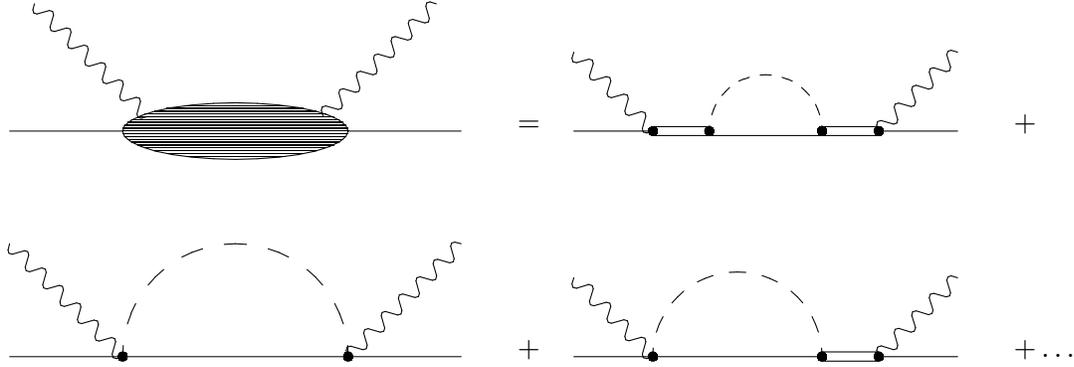

%\setcaptionmargin{5mm}
%\onelinecaptionsfalse
%\onelinecaptionstrue
\magnitude=2000
\GRAPH(hsize=15){
\mov(0,0){\lin(1,0)}%
\mov(3,0){\lin(1,0)}%
\mov(0,-2){\lin(4,0)}%
\mov(1.,-2){\Circle*(0.08)}%
\mov(3.,-2){\Circle*(0.08)}%
\mov(2,-2){\halfdashcirc(2.)[U]}%
\mov(1.,-2){\wavelin(-1.0,1.)}%
\mov(3,-2){\wavelin(1.0,1.)}%
\mov(2.,0){\Ellipse*(0.5)[4,1]}%
\mov(1.22,0.13){\wavelin(-1.0,1.)}%
\mov(2.78,0.13){\wavelin(1.0,1.)}%
\mov(4.5,0){=}%
\mov(4.5,-2){+}%
\mov(5.0,0){\lin(0.7,0)}%
\mov(5.7,0){\Circle*(0.08)}%
\mov(5.7,0){\wavelin(-0.7,0.7)}%
\mov(5.7,-0.04){\lin(0.5,0)}%
\mov(5.7,0.04){\lin(0.5,0)}%
\mov(6.2,0){\Circle*(0.08)}%
\mov(6.2,-0.04){\lin(1,0)}%
\mov(7.7,0){\Circle*(0.08)}%
\mov(7.7,0){\wavelin(0.7,0.7)}%
\mov(7.2,-0.04){\lin(0.5,0)}%
\mov(7.2,0.04){\lin(0.5,0)}%
\mov(7.2,0){\Circle*(0.08)}%
\mov(7.7,0){\lin(0.7,0)}%
\mov(6.7,0){\halfdashcirc(1.0)[U]}%
\mov(8.9,0){+}%
\mov(8.9,-2){$+\ldots$}%
\mov(5.0,-2){\lin(0.7,0)}%
\mov(5.7,-2){\Circle*(0.08)}%
\mov(5.7,-2){\wavelin(-0.7,0.7)}%
\mov(5.7,-2){\lin(1.5,0)}%
\mov(7.7,-2){\Circle*(0.08)}%
\mov(7.7,-2){\wavelin(0.7,0.7)}%
\mov(7.2,-1.96){\lin(0.5,0)}%
\mov(7.2,-2.04){\lin(0.5,0)}%
\mov(7.2,-2){\Circle*(0.08)}%
\mov(7.7,-2){\lin(0.7,0)}%
\mov(6.45,-2){\halfdashcirc(1.5)[U]}%
}
%\captionstyle{normal}
\caption{The proton polarizability correction in the resonance region.
On the Feynman diagram solid, double solid, wave and dashed lines correspond
to the nucleon, baryon resonance, photon and pion respectively.}
\end{figure*}

In the resonance region the proton polarizability contribution to the
Lamb shift is determined by the processes of a $\pi$-, $\eta$-meson production
on nucleons and the production of basic low-lying nucleon resonances.
Several amplitudes of such reactions are presented in Fig.2. To calculate
the contributions of separate resonances to the cross sections $\sigma_{T,L}$
in the isobar model we used the Breit-Wigner parameterization suggested
in Refs.\cite{Walker,Arndt,T1,T2,KAA,Bianchi,UIM1,Dong1,T3,Dong2,LPS,IGA}.
In the considered region of the variables $k^2$, $W$ the most contribution
is given by five resonances: $P_{33}(1232)$, $S_{11}(1535)$, $D_{13}(1520)$,
$P_{11}(1440)$, $F_{15}(1680)$. Accounting the resonance decays to the
$N\pi$- and $N\eta$-states we can express the absorption cross sections
$\sigma_{1/2}$ and $\sigma_{3/2}$ as follows:
\begin{equation}
\sigma_{1/2,3/2}=\left(\frac{k_R}{k}\right)^2\frac{W^2\Gamma_\gamma\Gamma_{R
\rightarrow N\pi}}{(W^2-M_R^2)^2+W^2\Gamma_{tot}^2}\frac{4m_p}{M_R\Gamma_R}
|A_{1/2,3/2}|^2,
\end{equation}
where ${\rm A_{1/2,3/2}}$ are transverse electromagnetic helicity amplitudes,
\begin{equation}
\Gamma_\gamma=\Gamma_R\left(\frac{k}{k_R}\right)^{j_1}\left(\frac{k_R^2+X^2}
{k^2+X^2}\right)^{j_2},~~X=0.3~GeV.
\end{equation}
The resonance parameters $\Gamma_R$, $M_R$, $j_1$, $j_2$, $\Gamma_{tot}$ are
taken from Refs.\cite{PDG,T2,T3,Dong1}. In accordance with Refs.\cite{T1,KAA,T3}
the parameterization of a one-pion decay width is
\begin{equation}
\Gamma_{R\rightarrow N\pi}(q)=\Gamma_R\frac{M_R}{M}\left(\frac{q}{q_R}\right)^3
\left(\frac{q_R^2+C^2}{q^2+C^2}\right)^2,~~C=0.3~GeV
\end{equation}
for the resonance $P_{33}(1232)$ and
\begin{equation}
\Gamma_{R\rightarrow N\pi}(q)=\Gamma_R\left(\frac{q}{q_R}\right)^{2l+1}
\left(\frac{q_R^2+\delta^2}{q^2+\delta^2}\right)^{l+1},
\end{equation}
for $D_{13}(1520)$, $P_{11}(1440)$, $F_{15}(1680)$. $l$ is the pion angular
momentum, $\delta^2=(M_R-m_p-m_\pi)^2+\Gamma_R^2/4$. Here $q$ $(k)$ and $q_R$ $(k_R)$
denote the c.m.s. pion (photon) momenta of resonances with the mass
$M$ and $M_R$ respectively. In the case of $S_{11}(1535)$ we take into account
$\pi N$ and $\eta N$ decay modes \cite{KAA,T3}:
\begin{equation}
\Gamma_{R\rightarrow N\pi,N\eta}=\frac{q_{\pi,\eta}}{q}b_{\pi,\eta}\Gamma_R
\frac{q_{\pi,\eta}^2+C_{\pi,\eta}^2}{q^2+C_{\pi,\eta}^2},
\end{equation}
where $b_{\pi,\eta}$ is the $\pi$, $\eta$ branching ratio. The cross section
$\sigma_L$ is determined by an expression similar to Eq.(15) where we must
change $A_{1/2,3/2}$ on the longitudinal amplitude $S_{1/2}$. The calculation
of helicity amplitudes $A_{1/2}$, $A_{3/2}$, $S_{1/2}$ as functions of
$Q^2$ was done on the basis of the oscillator quark model in
Refs.\cite{Dong2,Isgur,CL,Capstick,LBL,Warns}.

The two-pion decay modes of the higher nucleon resonances $S_{11}(1535)$,
$D_{13}(1520)$, $P_{11}(1440)$, $F_{15}(1680)$ are described phenomenologically
using the two-step process as in Ref.\cite{T3}. The high-lying nucleon resonance
$R$ can decay first into $N^\ast$ ($P_{33}$ or $P_{11}$) and a pion or
into a nucleon and $\rho$, $\sigma$ meson. Then the new resonances decay
into a nucleon and a pion or two pions:
\begin{equation}
R\rightarrow r+a=\Biggl\{{N^\ast+\pi\rightarrow
N+\pi+\pi,\atop \rho(\sigma)+N\rightarrow N+\pi+\pi.}
\end{equation}
The total decay width of such processes can be presented as a phase-space
weight integral over the mass distribution of the intermediate resonance
$r$ = $N^\ast, \rho, \sigma$ ($ a=\pi, N$):
\begin{equation}
\Gamma_{R\rightarrow
r+a}(W)=\frac{P_{2\pi}}{W}\int_0^{W-m_a}d\mu \cdot
p_f\frac{2}{\pi}
\frac{\mu^2\Gamma_{r,tot}(\mu)}{(\mu^2-m_r^2)^2+\mu^2\Gamma_{r,tot}^2(\mu)}
\frac{(M_R-m_2-2m_\pi)^2+C^2}{(W-m_2-2m_\pi)^2+C^2},
\end{equation}
where $C=0.3~GeV$. The factor $P_{2\pi}$ must be taken from the constraint
condition: $\Gamma_{R\to r+a}(W_R)$ coincides with the experimental data
in the resonance point. $p_f$  is the three-momentum of the resonance $r$
in the rest frame of $R$. $\Gamma_{r,tot}$ is the total width of the
resonance $r$. The decay width of the meson resonance in Eq.(21) is
parameterized similarly to that of the $P_{33}(1232)$:
\begin{equation}
\Gamma(\mu)=\Gamma_r\frac{m_r}{\mu}\left(\frac{q}{q_r}\right)^{2J_r+1}
\frac{q_r^2+\delta^2}{q^2+\delta^2},~~~\delta=0.3~GeV,
\end{equation}
where $m_r$ and $\mu$ are the mean mass and the actual mass of the meson
resonance, $q$ and $q_r$ are the pion three momenta in the rest frame
of the resonance with masses $\mu$ and $m_r$. $J_r$ and $\Gamma_r$ are the
spin and decay width of the resonance with the mass $m_r$.

Main nonresonant contribution to the cross sections $\sigma_{T,L}$
in the resonance region is determined by the Born terms constructed
on the basis of Lagrangians of $\gamma NN$, $\gamma\pi\pi$, $\pi NN$
interactions. Another part of nonresonant background comprises the
$t$- channel contributions of $\rho$, $\omega$ mesons obtained by means
of effective Lagrangians $\gamma \pi V$, $VNN$ interactions ($V=\rho,\omega$)
\cite{UIM1,UIM2}. In the unitary isobar model accounting the Born terms,
the vector meson, nucleon resonance contributions and the interference
terms we calculated the cross sections $\sigma_{T,L}$ by means of numerical
program MAID (http://www.kph-uni-mainz.de/MAID) in the resonance region
as the functions of two variables $W$ and $Q^2$. Then the structure function
$F_2(W,Q^2)$ can be presented as follows:
\begin{equation}
F_2(W,Q^2)=\frac{Q^2}{4\pi^2\alpha}\left(\sigma_T+\sigma_L\right)\frac{K\nu}
{(Q^2+\nu^2)},
\end{equation}
where $K$ is the flux factor of virtual photons. There are two definitions
for the quantity $K$: the Gilman definition $K_G=\sqrt{Q^2+\nu^2}$, and the Hand
definition $K_H=\nu(1-Q^2/2m_2\nu)=(W^2-m_2^2)/2m_2$ \cite{Close}.
Our results are presented in Fig.3 for the total cross section
$\sigma_{tot}(W,Q^2)=(\sigma_T+\sigma_L)$. It contains three clear peaks
corresponding to resonances $P_{33}(1232)$, $D_{13}(1520)$, $F_{15}(1680)$.
The photoabsorption cross section $\sigma_{tot}(W,Q^2=0)$ differs from
experimental data in the range $1.5\leq W\leq 2$ GeV obtained in
Refs.\cite{TAA,MM}. So, theoretical model must be improved by the account
of two-pion resonance decays as described in Eq.(21).

\begin{figure}[t!]
%\setcaptionmargin{5mm}
%\onelinecaptionsfalse
\includegraphics{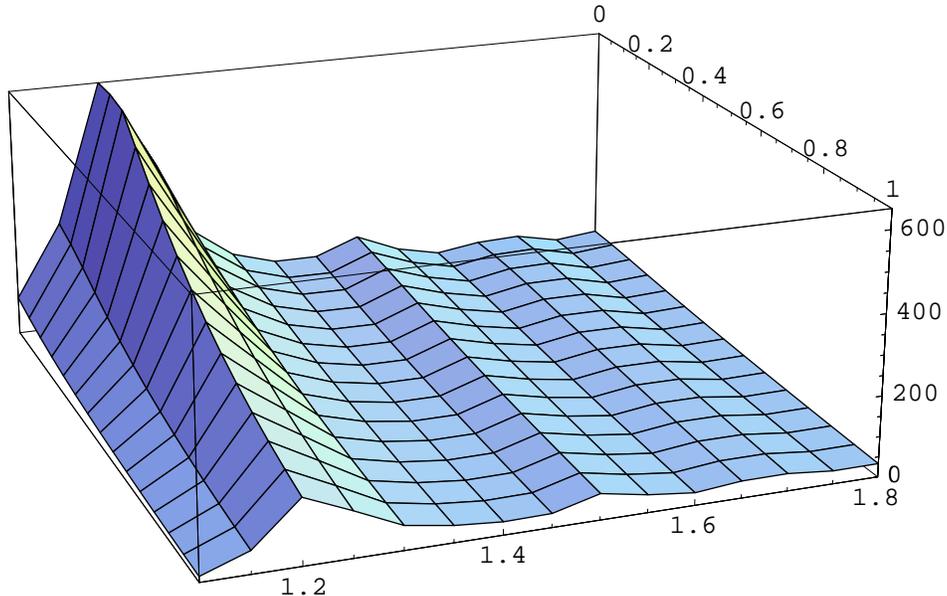}
%\captionstyle{normal}
\caption{Total photoabsorption cross section $\sigma_{tot}(W,Q^2)=
(\sigma_T+\sigma_L)$ in $\mu b$ as the function of variables $Q^2$
$(0\div 1)~GeV^2$ and $W$ $(1.1\div 1.8)~GeV$.}
\end{figure}

In the nonresonant region there exist several parameterizations
for the function $F_2(Q^2,W)$ \cite{BK} obtained on the basis of
experimental data on deep inelastic lepton-nucleon scattering.
The structure function $F_2$ can be expressed in terms of parton
distributions
\begin{equation}
F_2(x,Q^2)=\sum_i e_i^2xq_i(x,Q^2).
\end{equation}
So, to construct it in deep inelastic region we can use the
$Q^2$ evolution equations for the quark and gluon distributions
\cite{DGLAP}:
\begin{equation}
\frac{dq_i(x,Q^2)}{d\ln Q^2}=\frac{\alpha_s}{2\pi}\int_x^1\frac{dy}{y}
\left[q_i(y,Q^2)P_{qq}\left(\frac{x}{y}\right)+g(y,Q^2)P_{qg}\left(
\frac{x}{y}\right)\right],
\end{equation}
\begin{equation}
\frac{dg(x,Q^2)}{d\ln Q^2}=\frac{\alpha_s}{2\pi}\int_x^1\frac{dy}{y}
\left[\sum_iq_i(y,Q^2)P_{gq}\left(\frac{x}{y}\right)+g(y,Q^2)P_{gg}\left(
\frac{x}{y}\right)\right],
\end{equation}
where the sum is considered over all quarks and antiquarks.
$P_{qq}$, $P_{gq}$, $P_{qg}$, $P_{gg}$ are the quark-gluon splitting
functions \cite{LP}. Numerical solution of the integrodifferential evolution
equations (25), (26) by means of the method suggested in Ref.\cite{Kumano}
allows to obtain the parton distributions and the structure function
$F_2(x,Q^2)$ for different values
of a photon momentum squared $Q^2$. Corresponding numerical results are
in good agreement with the world experimental data. On the other hand, the
23 parameter model for the structure function $F_2(x,Q^2)$ based on
experimental data was proposed in Ref.\cite{AL}. Here it was expressed as
a sum of the Pomeron $F_2^{\cal P}$ and the Reggeon $F_2^{\cal R}$ term
contributions:
\begin{equation}
F_2(x,Q^2)=\frac{Q^2}{Q^2+m_0^2}\left[F_2^{\cal R}(x,Q^2)+F_2^{\cal P}(x,Q^2)
\right],
\end{equation}
\begin{equation}
F_2^{\cal R}(x,Q^2)=C_{\cal R}(t)x_{\cal R}^{a_{\cal R}(t)}(1-x)^{b_{\cal R}
(t)},~~~
F_2^{\cal P}(x,Q^2)=C_{\cal P}(t)x_{\cal P}^{a_{\cal P}(t)}(1-x)^{b_{\cal P}
(t)},
\end{equation}
where $x$ is the Bjorken variable,
\begin{equation}
\frac{1}{x_{\cal R}}=1+\frac{W^2-m_2^2}{Q^2+m_{\cal R}^2},~~~
\frac{1}{x_{\cal P}}=1+\frac{W^2-m_2^2}{Q^2+m_{\cal P}^2},
\end{equation}
More accurate numerical values of the model parameters are presented in
Ref.\cite{AL} (version 2 (2004)).

For the second structure function $R(Q^2,W)$ in nonresonant region
we used also the parameterization obtained with the aid of experimental
data \cite{Abe}. In the resonance region there are no experimental data
for the quantity $R(Q^2,W)$. In the most important part of the resonance
region $\sigma_L\ll \sigma_T$. So, to perform the numerical calculation
of the correction $\Delta E^{LS}_{pol}$ we supposed that $R(Q^2,W)\approx 0$.

\section{Numerical results}

In this paper we calculate the proton polarizability correction
to the Lamb shift of electronic and muonic hydrogen on the basis of
the isobar model describing the processes of low-energy scattering
of virtual photons on nucleons and the evolution equations for
the parton distributions. These two significant ingredients of the calculation
allow to construct the absorption cross sections for transversely and
longitudinally polarized photons by nucleons $\sigma_{T,L}$ and to express
the structure function $F_2(Q^2,W)$ (23), (27), which determines required
contribution (10). Numerical results are presented in Table 1.
We investigated contributions to the correction (10) which have numerical
value of order 1 $\mu eV$ for the $1S$ state in muonic hydrogen and
1 $Hz$ for the $1S$ state in electronic hydrogen. It is inferred from these
results that the basic contribution to the polarizability effect is given
by processes of the $\pi$-meson production on nucleons in the reaction
$\gamma^\ast+N\to\pi+N$ including the resonant reactions (2 line in Table 1).
In the isobar model we kept also processes of the $\eta$-meson production
on nucleons \cite{UIM2} (3 line in Table 1) and the production processes
of the $K$-mesons (4 line in Table 1). Declared accuracy of the calculation
calls for further consideration of the two-pion final states in the reaction
$\gamma^\ast+ N\to N+\pi+\pi$ (5 line in Table 1). Indeed, the comparison
of the total photoabsorption cross section, derived from the isobar model,
with experimental data shows that in the resonance region at
$2~GeV\geq W\geq 1.5~GeV$ theoretical photoabsorption cross section is less
than experimental one by the value 100 $\mu b$ approximately. Two-pion
final states in the resonant reactions of the form (20) are taken into
account in the construction of the cross sections $\sigma_{T,L}$
through the use of a two-stage model (21). Corresponding contribution
to the correction (10) is equal to (-6 $\mu eV$). But there exists nonresonant
contribution of two-pion final states to the total cross section
$\sigma_{tot}(Q^2,W)$. We included in Table 1 approximate estimate of this
contribution equal to (-6 $\mu eV$). It is based on the assumption that in the
whole range of variables $Q^2$, $W$: $0\leq Q^2\leq 1$, $1.5\leq W\leq 2$
determining the value of the correction (10) the total cross section
of virtual photoabsorption is less than experimental data by the value
of order 100 $\mu eV$ as for the cross section at $Q^2=0$.

\begin{table}[!ht]
%\setcaptionmargin{0mm}
%\onelinecaptionsfalse
%\captionstyle{flushleft}
\caption{Proton polarizability correction in the Lamb shift of electronic
and muonic hydrogen.}
\bigskip
\begin{tabular}{|c|c|c|c|c|}     \hline
Contribution of the reaction (13) & \multicolumn{2}{|c|}{$(e^-p^+)$~~~Hz} &
\multicolumn{2}{|c|}{$(\mu^- p^+)~~~\mu eV$} \\ \cline{2-3}\cline{4-5}
to the correction $\Delta E_{pol}^{LS}$ & $1S$ & $2S$ & $1S$ & $2S$  \\ \hline
Contribution of $N\pi$-states &- 86.81&-10.851 & -103.3&-12.91  \\  \hline
Contribution of $N\eta$-states &-0.02&-0.003  & -0.6&-0.08  \\  \hline
Contribution of $K$-mesons &-0.03&-0.004  &-1.1&-0.14  \\  \hline
Contribution of $N\pi\pi$-states &- 0.54&-0.068 &-12.0&-1.50   \\  \hline
Nonresonant contribution &-0.44&-0.055  & -12.0&-1.50 \\  \hline
Contribution of subtraction &0.99&0.124   & 18.4&2.30  \\
term           &   & & &  \\  \hline
Summary contribution  &-86.85&-10.86   & -110.6&-13.83  \\  \hline
\end{tabular}
\end{table}

There exists a number of theoretical uncertainties connected with quantities entering
in the correction (10). In the improved isobar model \cite{UIM1,UIM2} containing
14 resonances, we can omit theoretical error which arises due to the
insertion of other high-lying nucleon resonances. On our sight the main
theoretical error is closely related with the calculation of the helicity
amplitudes $A_{1/2}(Q^2)$, $A_{3/2}(Q^2)$, $S_{1/2}(Q^2)$ in the
quark model based on the oscillator potential \cite{Close}.
Only systematical experimental data for the helicity amplitudes
of the photoproduction on the nucleons $A_{1/2}(0)$, $A_{3/2}(0)$ are known
with sufficiently high accuracy to the present \cite{PDG}.
In the case of amplitudes for the electroproduction of the nucleon resonances
experimental data contain only their values at several points $Q^2$.
So, we have no consistent check for the predictions of the oscillator model.
Possible theoretical uncertainty connected with the calculation of amplitudes
$A_{1/2}(Q^2)$, $A_{3/2}(Q^2)$, $S_{1/2}(Q^2)$ with the account of relativistic
corrections can attain the value of order 10 $\%$. Then the theoretical error
for the correction (10) in the resonance region comprises 20 $\%$ from
the obtained value, that is $\pm 2~\mu eV$ for the energy level $2S$ in muonic
hydrogen. There is theoretical uncertainty in the contribution due to
two-pion nonresonant processes which is presented above. The error in this
case can constitute no less than 30 $\%$ from corresponding contribution
that is $\pm 2$ $\mu eV$. The essential part of the theoretical error
is connected with the subtraction term in the dispersion integral (4).
Indeed, the increase of the world average value for the proton magnetic
polarizability from $1.6\times 10^{-4}~fm^3$ to $1.9\times 10^{-4}~fm^3$
\cite{PDG} during last years leads to the decrease of the summary contribution
to the Lamb shift $(2P-2S)$ by 0.4 $\mu eV$. The error of $\beta_M$
indicated in Ref.\cite{PDG}, gives the uncertainty $\pm 0.6~\mu eV$ to the
theoretical result for the shift $\Delta E^{LS}_{pol}(2P-2S)$. The
obtained value of the proton polarizability contribution to the Lamb shift
$(2P-2S)$ in muonic hydrogen is equal to $(13.8\pm 2.9)~\mu eV$. It is
intermediate in the value between the result $(16\div 17)~\mu eV$
calculated in Refs.\cite{RR,FM1} and $12~\mu eV$, obtained in Ref.\cite{P1}.
Our theoretical uncertainty is slightly higher than in Refs.\cite{RR,FM1,P1},
because of the presence of a number of additional theoretical quantities
in the isobar model with the definite theoretical uncertainties.
In the case of electronic hydrogen the value of the proton polarizability
contribution to the Lamb shift is determined for the most part by the
process $\gamma^\ast+N\to\pi+N$ including the resonance contribution.
The obtained shift $(-87)~Hz$ of the $1S$ energy level in $(e^-p^+)$
is also in the agreement with the previously derived results
$(-72)~Hz$ in Ref.\cite{SPH}, $(-71)~Hz$ in Ref.\cite{IBK} and
$(-95)~Hz$ in Ref.\cite{RR}.

\begin{acknowledgments}
I am grateful to F.Kottmann for valuable information about the experiment
at PSI and R.N.Faustov for useful discussions. The work is performed
under the financial support of the Russian Foundation for Basic Research
(grant 04.02.16085) and the Program "Universities of Russia -
Fundamental Researches" (grant UR.01.02.367).
\end{acknowledgments}

\end{document}